\documentclass[12pt]{iopart}
\usepackage{times}
\usepackage{iopams}
\usepackage{bm}
\usepackage[dvips]{graphicx}

\begin{document}

\title[Spin multiplicity and entanglement swapping in radical ion recombinations]{Spin multiplicity and entanglement swapping in radical ion recombinations}
\author{L V Il'ichov${}^{~a,c}$ and S V Anishchik${}^{~b,c}$}
\address{${}^{a~}$Institute of Automation and Electrometry SB RAS, 630090 Novosibirsk, Russia, \\
${}^{b~}$Institute of Chemical Kinetics and Combustion SB
RAS, 630090 Novosibirsk, Russia, \\
 ${}^{c~}$Novosibirsk State University, 630090 Novosibirsk, Russia.}
\ead{leonid@iae.nsk.su and svan@kinetics.nsc.ru}

\begin{abstract}
{We address the problem of relative frequencies of singlet and
triplet recombinations in a multiparticle system, which consists
of spin-correlated radical ion pairs.  The nonlocal swapping of
spin correlations due to cross-recombinations is taken into
account. It is shown that this swapping does not contribute to
singlet and triplet recombination frequencies in the absence of
spin evolution in the correlated pairs.}
\end{abstract}
%%% PACS numbers

\section{Introduction}

\begin{figure}
\centering\includegraphics[width =
0.3\textwidth]{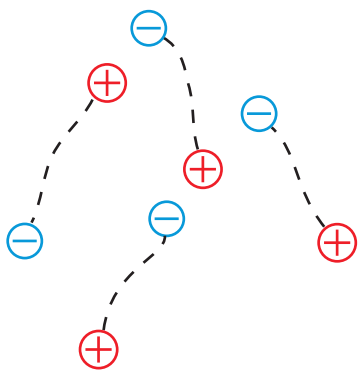}\caption{(Color online) The set of
geminal pairs (connected with dotted lines). All the pairs are in
their singlet spin state.} \label{f:1}
\end{figure}

The track created by a swift particle in a medium has a rather
complicated structure, which depends on the type and energy of the
particle \cite{Mozumder69}. This is due to specific processes of
particle energy loss in the course of its collisions with
molecules in the medium. Several radical ion pairs can be born in
the same spatial region. These particles can then undergo either
geminal recombination (within the initial pair) or
cross-recombination. All geminal pairs are initially in their
singlet spin state (\Fref{f:1}). The multiplicity of the pair can
be changed by spin interactions with nuclei in the molecules and
with external magnetic fields. The recombination probability per
se is not affected by the multiplicity of the pair, but only
singlet recombination normally leads to luminescence. Hence,
external magnetic fields can influence the intensity of the
recombination luminescence \cite{Kniga}. Investigation of this
influence can provide important information on the structure of
the radical ions, chemical processes which the radicals are
involved in, and the structure of the radiation track
\cite{Usov,Borovkov}.

It is generally believed that the intensity of luminescence
arising from cross-recombinations is not affected by magnetic
field, since the spins of the particles are not initially
correlated and, consequently, the probability of singlet
recombination is universally equal to $1/4$. This is supported by
experiments on magnetic field effects in recombination
luminescence induced by various sources of ionizing radiation
\cite{Brocklehurst97}. The magnitude of the effect tends to
decrease as the track density increases and, consequently, the
fraction of cross-recombinations grows higher. Nevertheless,
 the question on the exact absence of magnetic effects in 
cross-recombinations is still open.
Some doubts are brought by magnetic field effect on alkane
solutions irradiated with helium ions of 20 MeV \cite{LaVerne69}.
 Thus created tracks are extremely dense, which leads to a great 
number of cross-recombinations. Still a small magnetic effect has really
been detected.

The magnetic effect in cross-recombinations can hypothetically be
caused by a non-local quantum phenomenon related to the effect
known in quantum information theory as entanglement swapping
\cite{b2}. Brocklehurst was the first (even before the appearance
of \cite{b2}) to pay attention to this phenomenon \cite{b3}. Its
essence is as follows. Let there be no spin evolution in the
geminal pairs. Then each geminal pair is in its singlet spin state
for the period of time up to cross-recombination of any of its
member. After the singlet cross-recombination of the fragments
$\{2,3\}$ of two singlet geminal pairs $\{1,2\}$ and $\{3,4\}$ the
remaining pair of fragments $\{1,4\}$ appears to be in the singlet
state. This follows from the conservation of the total zero spin
of the four particles. Note that the \textsl{a priori} (before the
recombination of the radicals 2 and 3) spin state of the pair
$\{1,4\}$ was maximally mixed and, naturally, demonstrated no spin
correlation. In such a manner the entanglement contained in the
singlet state has been swapped inside the quadruple $\{1,2,3,4\}$
from the couples $\{1,2\}$ and $\{3,4\}$ to the new ones $\{2,3\}$
and $\{1,4\}$. The first of these new pairs is the product of
recombination and quits the subsequent consideration. The pair
$\{1,4\}$ continues its life in the medium. It can take part in a
new act of entanglement swapping due to cross-recombination.
Otherwise upon meeting of its fragments it will disappear in
singlet recombination with unity probability. It is worth noting
for the following that this unity probability is conditioned by
the previous recombination of the pair $\{2,3\}$.

Now let us suppose that the event of triplet recombination of the
pair $\{2,3\}$ took place and we know nothing about the
orientation and alignment of the product. This is the situation of
its isotropic triplet state. The same reasoning brings us to the
conclusion that the pair $\{1,4\}$ is also in the isotropic
triplet state. This state is not entangled and only contains
classical spin correlations, which are the result of swapping and
transformation of the former singletness of the pairs $\{1,2\}$
and $\{3,4\}$. The meetings and recombinations of the fragments of
the triplet pairs induce further spin correlation swapping (with
the accompanying transformation) into the remaining pair, etc.

Note that a non-local chemical reaction takes place when a
spatially-separated pair of radicals in a definite spin state
appears as a result of the distant recombination. It was shown in
\cite{Ilichev08} that the swapping process significantly affects
the spatial correlations of the recombination events.

The role of swapping of interparticle spin correlations in the
relative frequencies of singlet and triplet recombinations seems
to be important. Suppose that the swapping does affect the
relative frequencies. By this way  spin evolution in magnetic
fields can modify the frequencies. We shall show that under
certain conditions the swapping process does not manifest itself
in the triplet and singlet recombination frequencies.

\section{Model}

We assume that no spin evolution takes place between the events of
creation and recombinations. We then introduce the spin
statistical operator $\hat{\varrho}_{a,b}$ for the pair of
radicals with numbers $a$ and $b$. This operator acts in the
tensor product $\mathcal{H}_{a}\otimes\mathcal{H}_{b}$ of spin
Hilbert states of the particles. We also need the following
enumerable set of statistical operators labeled with the parameter
$n = 0,1,\ldots$:
\begin{equation}
\hat{\varrho}^{(n)}_{a,b} =
\frac{1}{4}\bigg[\hat{\sigma}_{0}(a)\otimes\hat{\sigma}_{0}(b) -
\bigg(-\frac{1}{3}\bigg)^{n}\hat{\sigma}_{k}(a)\otimes\hat{\sigma}_{k}(b)\bigg].
\label{1}
\end{equation}
Here $\hat{\sigma}_{0}$ is the unit matrix $2\times2$;
$\hat{\sigma}_{k}$ are Pauli matrices , $k$ is the spatial axis
index ( x, y or z; the summation over repeated indices is
assumed). One can readily check that $\hat{\varrho}^{(0)}$ is the
singlet state and $\hat{\varrho}^{(1)}$ is the isotropic triplet
state. The set $\{\hat{\varrho}^{(n)}\}_{n=0}^{\infty}$ includes
all the pair spin states which can appear in the course of
recombinations and swapping process. To prove this we introduce
the swapping map $S_{2,3}$ from the set of quadruple spin states
to the set of the states of the couple $\{1,4\}$ under the
condition of singlet recombination of radicals 2 and 3, which
initially belong to correlated pairs $\{1,2\}$ and $\{3,4\}$
\cite{Il01}:
\begin{equation}
\mathcal{S}_{2,3}:  \hat{\varrho}^{(pre)}_{\,1,2,3,4} =
\hat{\varrho}_{\,1,2}\otimes \hat{\varrho}_{\,3,4} \mapsto
\hat{\varrho}^{(post)}_{1,4}. \label{2}
\end{equation}
\begin{figure}
\centering\includegraphics[width =
0.45\textwidth]{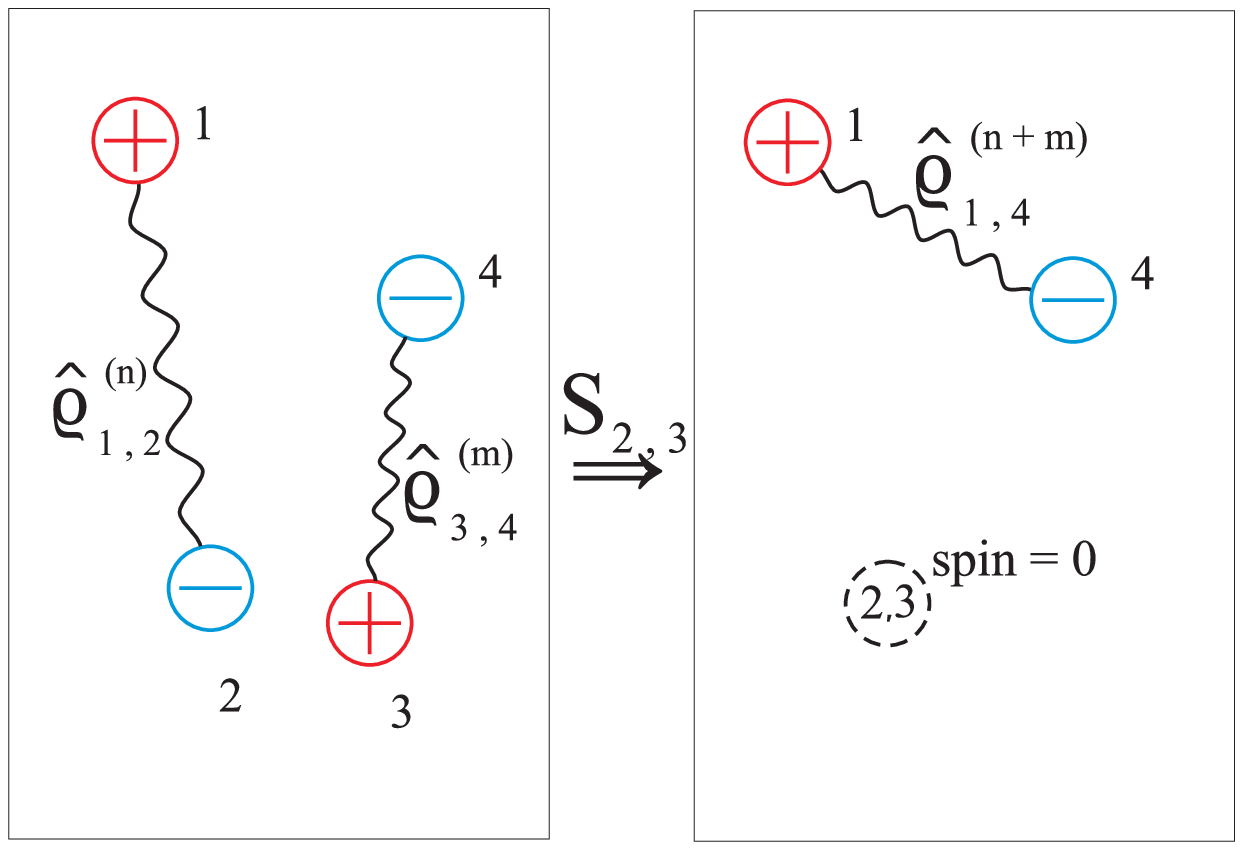}\caption{(Color online) The swapping of
spin correlations in the course of \emph{singlet} recombination of
particles 2 and 3. The dotted circle represents the product of
recombination.} \label{f:2}
%\end{figure}
%\begin{figure}
\centering\includegraphics[width =
0.45\textwidth]{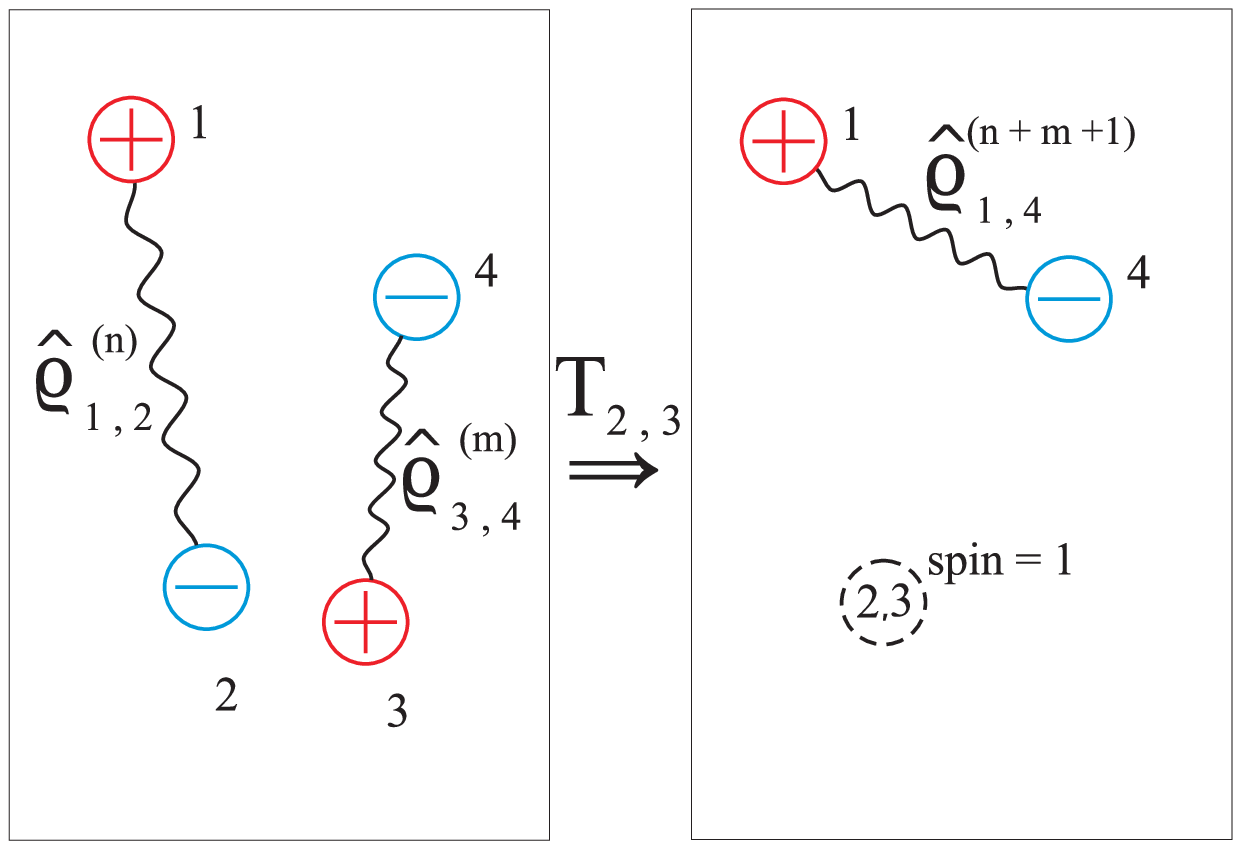}\caption{(Color online) The swapping of
spin correlations in the course of \emph{triplet} recombination of
particles 2 and 3.} \label{f:3}
\end{figure}
The event of singlet recombination is equivalent to the positive
result of an experiment checking the singlet status of the pair
$\{2,3\}$. Hence, we have
\begin{equation}
\hat{\varrho}^{(post)}_{\,1,4} =
\frac{Tr_{\,2,3}\,\hat{\varrho}^{(0)}_{\,2,3}\,\hat{\varrho}^{(pre)}_{\,1,2,3,4}}
{Tr_{\,1,2,3,4}\,\hat{\varrho}^{(0)}_{\,2,3}\,\hat{\varrho}^{(pre)}_{\,1,2,3,4}}.
\label{3}
\end{equation}
It is easy to show that the map $\mathcal{S}$ has the following
simple realization in the set
$\{\hat{\varrho}^{(n)}\}_{n=0}^{\infty}$ (\Fref{f:2}):
\begin{equation}
\mathcal{S}_{\,2,3}: \hat{\varrho}^{(n)}_{\,1,2}\otimes
\hat{\varrho}^{(m)}_{\,3,4} \mapsto \hat{\varrho}^{(n+m)}_{\,1,4}.
\label{4}
\end{equation}

In a similar way we introduce the map for the case of (isotropic)
triplet recombination of the pair $\{2,3\}$:
\begin{equation} \label{5}
\mathcal{T}_{\,2,3}: \hat{\varrho}^{(pre)}_{\,1,2,3,4} =
\hat{\varrho}_{\,1,2}\otimes \hat{\varrho}_{\,3,4} \mapsto
\frac{Tr_{\,2,3}[\hat{\sigma}_{0}(2)\otimes\hat{\sigma}_{0}(3) -
\hat{\varrho}^{(0)}_{\,2,3}]\hat{\varrho}^{(pre)}_{\,1,2,3,4}}
{Tr_{\,1,2,3,4}[\hat{\sigma}_{0}(2)\otimes\hat{\sigma}_{0}(3) -
\hat{\varrho}^{(0)}_{\,2,3}]\hat{\varrho}^{(pre)}_{\,1,2,3,4}}.
\end{equation}
In the set $\{\hat{\varrho}^{(n)}\}_{n=0}^{\infty}$ one gets
(\Fref{f:3}):
\begin{equation}
\mathcal{T}_{2,3}: \hat{\varrho}^{(n)}_{\,1,2}\otimes
\hat{\varrho}^{(m)}_{\,3,4} \mapsto
\hat{\varrho}^{(n+m+1)}_{\,1,4}. \label{6}
\end{equation}
We see that if only singlet and triplet pairs are created in the
medium by irradiation, the swapping maps $\mathcal{S}$ and
$\mathcal{T}$ generate the entire set
$\{\hat{\varrho}^{(n)}\}_{n=0}^{\infty}$ of the spin states.

Let $f^{(n)}(\mathbf{r}_1,\mathbf{r}_2)$ be the density of pairs
in the spin state $\hat{\varrho}^{(n)}$ and with locations of
radicals near the points $\mathbf{r}_1$ and $\mathbf{r}_2$. The
first partner in every pair is the radical of the type "$+$", and
the second -- of the type "$-$". Recombinations take place only
between partners of different types. We also introduce the radical
spatial densities $g_{+}(\mathbf{r})
=\sum_{n=0}^{\infty}\,\int\,f^{(n)}(\mathbf{r},
\mathbf{r}')\,d^{3}\mathbf{r}'$ and $g_{-}(\mathbf{r}) =
\sum_{n=0}^{\infty}\,\int\,f^{(n)}(\mathbf{r}',
\mathbf{r})\,d^{3}\mathbf{r}'$, which can readily be obtained from
$f^{(n)}(\mathbf{r}_1,\mathbf{r}_2)$.
These notions let us introduce the following important operator
$\langle\hat{\varrho}\,(\mathbf{r}_1,\mathbf{r}_2)\rangle$, which
is the average spin state of a "$+-$"-pair, localized at the
points $\mathbf{r}_1$ and $\mathbf{r}_2$:
\begin{equation} \label{7}
\langle\hat{\varrho}\,(\mathbf{r}_1,\mathbf{r}_2)\rangle =
\frac{1}{4}\bigg(1 -
\sum_{n}\,\frac{f^{(n)}(\mathbf{r}_1,\mathbf{r}_2)}
{g_{+}(\mathbf{r}_1)g_{-}(\mathbf{r}_2)}\bigg)
\hat{\sigma}_{0}\otimes\hat{\sigma}_{0}\; +
\sum_{n}\,\frac{f^{(n)}(\mathbf{r}_1,\mathbf{r}_2)}
{g_{+}(\mathbf{r}_1)g_{-}(\mathbf{r}_2)}\hat{\varrho}^{(n)}.
\end{equation}
The first term is the maximally mixed state with the proper weight
$\hat{\sigma}_{0}\otimes\hat{\sigma}_{0}/4$. This is the case when
the pair was neither geminal, nor created in the swapping process.
The second term reflects the contribution of situations when the
pair appears to be correlated due to some sequence of swapping
acts.

The expression for
$\langle\hat{\varrho}\,(\mathbf{r}_1,\mathbf{r}_2)\rangle$ has the
following simple form:
\begin{equation}
\langle\hat{\varrho}\,(\mathbf{r}_1,\mathbf{r}_2)\rangle =
\frac{1}{4}\bigg[\hat{\sigma}_{0}\otimes\hat{\sigma}_{0} -
\xi(\mathbf{r}_1,\mathbf{r}_2)\hat{\sigma}_{k}\otimes\hat{\sigma}_{k}\bigg],
\label{8}
\end{equation}
where
\begin{equation}
\xi(\mathbf{r}_1,\mathbf{r}_2) =
\sum_{n=0}^{\infty}\,\bigg(-\frac{1}{3}\bigg)^{n}\frac{f^{(n)}(\mathbf{r}_1,\mathbf{r}_2)}
{g_{+}(\mathbf{r}_1)g_{-}(\mathbf{r}_2)}. \label{9}
\end{equation}
If we average the parameter $\xi(\mathbf{r},\mathbf{r})$ over the
volume of the system, we get the weights of singlet and triplet
states of every meeting pair and, consequently, the relative
frequencies of singlet and triplet recombinations.

Evaluation of the parameter $\xi(\mathbf{r},\mathbf{r})$ requires
an explicit form of kinetic equation for
$f^{(n)}(\mathbf{r}_{1},\mathbf{r}_{2})$. We take the following
model system:
\begin{equation}
\partial_{t}f^{(n)}(\mathbf{r}_1,\mathbf{r}_2) =
-\kappa (g_{-}(\mathbf{r}_1) +
g_{+}(\mathbf{r}_2))f^{(n)}(\mathbf{r}_1,\mathbf{r}_2)\; +
\label{10}
\end{equation}
$$
+
\frac{\kappa}{4}\sum_{m=0}^{n}\int\,f^{(m)}(\mathbf{r}_1,\mathbf{r})
f^{(n-m)}(\mathbf{r},\mathbf{r}_2)d^{3}\mathbf{r}\; +
$$
$$
+
\frac{3\kappa}{4}\sum_{m=0}^{n-1}\int\,f^{(m)}(\mathbf{r}_1,\mathbf{r})
f^{(n-m-1)}(\mathbf{r},\mathbf{r}_2)d^{3}\mathbf{r}\; +
$$
$$
+ (D_{+}\Delta_{\mathbf{r}_1} +
D_{-}\Delta_{\mathbf{r}_2})f^{(n)}(\mathbf{r}_1,\mathbf{r}_2)\; +
\gamma^{(n)}w(|\mathbf{r}_1 - \mathbf{r}_2|).
$$
The first term in the r.h.s. is the loss of pairs because of
recombination of one of its fragments; $\kappa$ is the
recombination rate. The second and third terms stand for the
creation of a new pair due to swapping act caused by singlet and
triplet recombination, respectively (the recombination takes place
at the point $\mathbf{r}$). We assume that the cross-sections of
the singlet and triplet recombinations are equal (as in the case
of charged radicals). Under these conditions the factors 1/4 and
3/4, with which the recombination constant appears in these terms,
are stipulated by the weights of singlet and triplet states in the
maximally mixed spin state (this is the state of the meeting pair
of radicals -- fragments of different correlated pairs). The
summation over $m$ reflects the laws (\ref{4}) and (\ref{6}) of
spin state transformation in the course of swapping. The last line
contains the diffusion terms and the gain term due to ionization;
the parameter $\gamma^{(n)}$ is the generation rate (we assume
that only $\gamma^{(0)}$ and $\gamma^{(1)}$) are non-zero); the
form-factor $w(r)$ gives the distribution of the relative
positions of the created radicals.

\section{Results and Discussion}

Under the homogeneous spatial conditions the densities $g_{+}$ and
$g_{-}$ are identical and do not depend on coordinates. At the
same time the pair density is a function of the distance between
the constituents. Upon multiplication of (\ref{10}) by
$(-1/3)^{n}g^{-2}$ and summation over $n$ one gets
\begin{equation} \label{11}
\partial_{t}\xi(r) =
-2\kappa g\xi(r) + (D_{+} + D_{-})\Delta_{\mathbf{r}}\xi(r)\;
+\frac{\gamma^{(0)} - \gamma^{(1)}/3}{g^{2}}w(r).
\end{equation}
The main point is the annihilation of the terms in (\ref{10})
responsible for the swapping process. Hence this process makes no
contribution to the evolution of $\xi(r)$. This means that under
the absence of intermediate spin evolution the swapping does not
affect the relative frequency of singlet and triplet
recombinations. Note that the validity of this conclusion does not
depend on the form of the migration terms in the kinetic equation
(\ref{10}) (the diffusion terms we used are not the best possible
approximation).

As has already been mentioned, the parameter $\xi(r)$ for $r =0$
lets one evaluate the ratio of the frequencies of the singlet
($\nu_{S}$) and triplet ($\nu_{T}$) recombinations:
\begin{equation}
\frac{\nu_{S}}{\nu_{T}} = \frac{1 + 3\xi(0)}{3 - 3\xi(0)}.
\label{12}
\end{equation}
It should be stressed that in the ratio (\ref{12}) we take into
account all the recombinations -- between uncorrelated radicals as
well as between the fragments of a correlated pair. On the
contrary, only the recombinations of the first type should be
considered in the equation (\ref{10}).

\section{Conclusions}

The main result of the work is the conclusion that the swapping
process in the absence of spin evolution does not affect the
relative frequencies of singlet and triplet recombinations. For
instance, there is no spin evolution in the case of zero hyperfine
interaction between electrons and nuclei in radicals and equal
$g$-factors for radical cation and anion. This result may be
considered as an argument in support of the present approaches to
the evaluations of magnetic field effects, neglecting any possible
magnetic influence on cross-recombinational luminescence.
Nevertheless, a more accurate analysis with an explicit account of
spin evolution and temporal dependencies of singlet and triplet
recombination probabilities is required for description of a
pulsed experiment.

\ack

Financial support of RFBR (projects 07-02-00954 and
05-03-32801) is gratefully acknowledged.\\


\section*{References}

\begin{thebibliography}{99}

\bibitem{Mozumder69}
Mozumder A  1969  {\em Advances in Radiation Chemistry} (NY:
Wiley) {\bf 1} 1

\bibitem{Kniga}
Salikhov K M, Molin Yu N, Sagdeev RZ and Buchachenko A L 1984 {\it
Spin polarization and magnetic effects in radical reactions}
(Amsterdam: Elsevier)

\bibitem{Usov}
Anishchik S V, Usov O M, Anisimov O A and Molin Yu N 1998 {\it
Radiat. Phys. Chem.} {\bf 51} 31

\bibitem{Borovkov}
Borovkov V I and Velizhanin K A 2007 {\it Radiat. Phys. Chem.}
{\bf 76} 988

\bibitem{Brocklehurst97}
Brocklehurst B 1997 {\it Radiat. Phys. Chem.} {\bf 50} 213

\bibitem{LaVerne69}

La Verne J A and Brocklehurst B 1996 {\it Radiat. Phys. Chem.}
{\bf 47} 71


\bibitem{b2}
Zukovski M, Zeilinger A, Horne M A and Ekert A K  1993 {\it Phys.
Rev. Lett.} {\bf 71} 4287

\bibitem{b3}
Brocklehurst B  1985 {\it Int. Rev. Phys. Chem.} {\bf 4} 279

\bibitem{Ilichev08}

Il'ichev L V and Anishchik S V 2008  {\it JETP Lett.} {\bf 87} 45

\bibitem{Il01}
Il'ichev L V 2001 {\it Theor. Math. Phys.} {\bf 127} 549








\end{thebibliography}
\end{document}